\documentclass[11pt,a4paper]{article}
\pdfoutput=1
\usepackage{jheppub}
\usepackage[T1]{fontenc}
\usepackage{graphicx}
\usepackage{epstopdf}
\usepackage{slashed}
\usepackage{tikz}
\usepackage{amsmath}
\input epsf
\usepackage{epsfig}
\usepackage{amssymb}
\usepackage{graphics}
\usepackage[active]{srcltx}
\usepackage{graphicx}
\usepackage{subfig}
\usepackage{appendix}

\newcommand{\bea}{\begin{eqnarray}}
\newcommand{\eea}{\end{eqnarray}}
\newcommand{\bean}{\begin{eqnarray*}}
\newcommand{\eean}{\end{eqnarray*}}

\def\eref#1{(\ref{#1})}

\def\Label#1{\label{#1}%
  \smash{\hbox to0pt{\raise1ex\hbox{\tiny[#1]}\hss}}}

\def\spaa #1{\langle #1\rangle}
\def\spbb #1{[#1]}
\def\spab #1{\langle #1]}

\title{Boundary Contributions of On-shell Recursion Relations With Multiple-line Deformation}

\author{Chang Hu,}
\author{Xiao-Di Li,}
\author{Yi Li}

\affiliation{Zhejiang Institute of Modern Physics, Department of Physics, Zhejiang University,\\No. 38 Zheda Road, Hangzhou310027, P.R. China}

\emailAdd{liyiphysics@zju.edu.cn}
\emailAdd{lixiaodi@zju.edu.cn}
\emailAdd{alberthu@zju.edu.cn}

\abstract{On-shell recursion relation has been recognized as a powerful tool for calculating tree level amplitudes in quantum field theory, but it doesn't work well when the residue of the deformed amplitude $\hat{A}(z)$ doesn't vanish at infinity of $z$. However, in such  situation, we still can get the right amplitude by computing the boundary contribution explicitly. In \cite{ArkaniHamed:2008yf}, background field method was first used to analyze the boundary behaviors of amplitudes with two deformed external lines in different theories. The same method has also been generalized to calculate the explicit boundary operators of some amplitudes with BCFW-like deformation in \cite{Jin:2015pua}.
In this paper, we will take a step further to generalize the method into the case
of multiple-line deformation, and to show how the boundary behaviors (even the boundary
contributions) can be extracted in the method. }

\keywords{Scattering amplitude, On-shell recursion relation, Multiple-line deformation, Background field method}

\begin{document}

\maketitle
\flushbottom

\section{Introduction}
Recent decades have witnessed the prosperity in the area of
scattering amplitudes, including the discovery of new mathematical
structures, the new formalisms of scattering amplitudes, and even
more important, the new methods for calculating scattering amplitude
more efficiently. Among these methods the on-shell recursion
relations, pioneered by the BCFW recursion relations
\cite{Britto:2004ap,Britto:2005fq}, have been proved to be very
powerful tools, which can be used to construct higher-point
tree-level amplitudes from lower-point ones. After the idea of
deforming two external momenta to capture the analytic structures of
tree-level amplitudes is introduced, quickly the deformation of
multiple lines and even all lines are used in
\cite{Risager:2005vk,Cohen:2010mi,Cheung:2015cba} to discuss
on-shell constructibility or others aspects of amplitudes.

The on-shell recursion relations are based on the Cauchy's theorem.
It says that under an appropriate deformation of a subset of $n$
momenta, $p_i \rightarrow \hat{p}_i(z)=p_i+zq_i$, with $z$ being a
complex parameter, the residue of $\hat{A}_n(z)/ z$ at $z=0$, which
is nothing but the $n$-point physical amplitude $A_n$, equals to the
minus of the sum of all other residues. We divide the latter into
two parts as $A_n=-\sum_{z_I} \text{Res}_{z=z_I}\hat{A}_n(z)/z -
B_n$. Here $\text{Res}_{z=z_I}\hat{A}_n(z)/z$ is a residue at finite
$z_I$ and factorizes into $\sum_{z_I} \hat{A}_L(z_I) \frac{1}{P^2_I}
\hat{A}_{R}(z_I)$ with $\hat{A}_L$ and $\hat{A}_R$ being lower-point
amplitudes, while $B_n$ is the residue at $z=\infty$ and doesn't
have the similar factorization, called the \textit{boundary
contribution} (or boundary term). Then we can see that if the
boundary contribution vanishes, the on-shell recursion relations
provide an efficient method for calculating $n$-point amplitude by
lower-point amplitudes, but these relations will meet problems when
$B_n$ doesn't vanish, which means $\hat{A}_n(z)$ doesn't vanish at
$z=\infty$.\footnote{It's equivalent to say $\hat{A}_n(z)=
\sum_{n=0}^{\infty} a_n z^n $, when we expand it around the infinity
point.}

Since the boundary term $B_n$ is vital for the on-shell recursion
relation,  many methods were proposed to deal with it. The first
step is to determine for which theories the boundary contributions
vanish in the on-shell recursion relations In
\cite{ArkaniHamed:2008yf,Cheung:2008dn}, the authors demonstrated
that the deformed amplitudes of a wide variety of theories vanish at
infinity by splitting the scattering process into a hard part and a
soft background. However, there are also some theories or some cases
where the boundary contributions don't vanish. Then in
\cite{Benincasa:2007xk,Boels:2010mj}, the authors choose to
introduce auxiliary fields to eliminate the boundary term. However
if the boundary term does exist, we can also try to separate $B_n$
from others and compute it explicit. Then in
\cite{Feng:2009ei,Feng:2010ku,Feng:2011twa} the authors try to
isolate the boundary term by analyzing the properties of Feynman
diagrams. And in \cite{Zhou:2014yaa}, collecting factorization
limits of all physical poles is applied to find the boundary
contribution. Another progress in this direction was made in
\cite{Benincasa:2011kn,Benincasa:2011pg,Feng:2011jxa}, where the
idea of expressing boundary terms as roots of amplitudes are
introduced, although it's not very practical. Then in
\cite{Feng:2014pia,Feng:2015qna,Jin:2014qya,Jin:2015pua}, multiple
steps of BCFW-like deformation were used to calculate the boundary
contribution step by step until getting the final results.

However, all these methods can only be applied to limited types of
theories, and then a more general method is needed. Hence, we want
to develop a general method which is applicable for broader
theories. In \cite{ArkaniHamed:2008yf,Cheung:2008dn,Cheung:2015cba},
the background field method was proved to be a good method to
analyze the boundary behavior of deformed amplitudes, then in
\cite{Jin:2015pua} the authors used the background field method to
calculate the boundary term explicitly in the case of two momenta
being deformed. All these guide us that we can generalize this
background field method into more general cases of multiple-leg
deformation for better efficiency. Although the same idea has been
exploited in \cite{ArkaniHamed:2008yf,Cheung:2008dn,Cheung:2015cba},
and especially in \cite{Cheung:2015cba} the boundary behavior of
deformed amplitudes of a lot of different theories in four dimension
were analyzed, we should emphasize that our objective is to compute
the boundary term explicitly rather than just analyzing the boundary
behavior, and our discussions are in general dimension.

Now we roughly explain the basic idea. First when
$z\rightarrow\infty$, $m$  deformed momenta $\hat{p}_i(z)\sim zq_i$
are much larger than other undeformed momenta $p_j$, so the original
scattering process can be regarded as the process of some hard
particles $\hat{p}_i(z)$ scattering in the soft background of other
soft particles $p_j$. From this point of view, the $z$-dependence of
$m$-deformed $n$-point amplitude $\widetilde{B}_m(z)$\footnote{The
$z$-denpendent part of $m$-deformed $n$-point amplitude
$\widetilde{B}_m(z)$ is not exactly equal to boundary term $B_n$,
since $\lim_{z\rightarrow
\infty}\hat{A}_n(z)\approx\widetilde{B}_m(z)$, and
$B_n=\text{Res}_{z=\infty}\hat{A}_n(z)/z\approx\text{Res}_{z=\infty}\widetilde{B}_m(z)/z
$.} is nothing, but the $m$-point scattering amplitudes in the
nontrivial soft background, which can be calculated by the
corresponding Feynman diagrams. With such an understanding, the
technical difficulty  becomes the reading of the Feynman rules,
including the interactive vertexes and the nontrivial propagators,
which will be discussed carefully in the paper.

The structure of the paper is following. In section \ref{boundary},
we briefly introduce the method of background field to calculate the
boundary term. In section \ref{real}, we begin to consider the
simplest example of real scalar theory. Then in section
\ref{yukawa}, the boundary contribution in Yukawa theory is computed
and some results of examples are represented. In section \ref{YM},
we give the general discussions of Yang-Mills theory, then apply it
to some explicit examples. In section \ref{conclusion}, we give the
conclusion.

%%%%%%%%%%%%%%%%%%%%%%%%%%%%%%
\section{Boundary Term of On-shell Recursion Relation} \label{boundary}
%%%%%%%%%%%%%%%%%%%%%%%%%%%%%%%%

In this paper we will follow the method proposed in
\cite{Jin:2015pua}, which will be briefly reviewed  in this section.
In \cite{Jin:2015pua}, the authors considered a $n$-point
correlation function with two deformed momenta given
by\footnote{Here we consider the $n$-point correlation function in
the momentum space. Under the LSZ reduction, each field $\Phi$ in
$x_i$ of the correlation function $\langle \Phi(x_1)\Phi(x_2)\cdots
\Phi(x_n) \rangle$ in position space is associated with an external
momentum $p_i$ in amplitude. So if we make the two-line deformation
$p_1\rightarrow \hat{p}_1(z), p_n\rightarrow \hat{p}_n(z)$ for an
amplitude $A_n$, then the corresponding fields $\Phi(x_1),
\Phi(x_n)$ in correlation function are assigned $\hat{p}_1$ and
$\hat{p}_n$ by LSZ reduction. }
\begin{align}
\hat{G}^{(2)}_n(z)=\langle \Phi(\hat{p}_1(z))\Phi(p_2)\cdots \Phi(\hat{p}_n(z)) \rangle
=\int D\Phi e^{i S[\Phi]} \Phi(\hat{p}_1(z))\Phi(p_2)\cdots \Phi(\hat{p}_n(z)).
\end{align}
After splitting the field $\Phi$ into a high energy (or hard)  part
$\Phi^{\Lambda}$ and a soft part $\Phi$ (still denoted by $\Phi$)
according to the energy scale $\Lambda\sim |zq_i|\gg p_j$, and
expanding the action, then the leading contribution is
\begin{align}
  \hat{G}^{(2)}_n(z)=\int D\Phi e^{i S[\Phi]} \hat{G}_2(z) \Phi_2(p_2)\cdots \Phi_{n-1}(p_{n-1}),
\end{align}
with
\begin{equation}
  \hat{G}_2(z)=\int \mathcal{D}\Phi^{\Lambda}\text{exp}(iS_2^{\Lambda}[\Phi^{\Lambda},\Phi]) \Phi_1^{\Lambda}\Phi_n^{\Lambda},
\end{equation}
where $S_2^{\Lambda}[\Phi^{\Lambda},\Phi]$ is the sum of terms
quadratic in $\Phi^{\Lambda}$ in the expansion of the
action.\footnote{The terms linear in $\Phi^{\Lambda}$ vanish because
of equation of motion, and in the case of two hard particles we only
need to consider quadratic terms and ignoring higher terms.} After
applying the LSZ reduction to the $n$-point correlation  function
with two hard fields, we can get the large $z$-dependent part of the
deformed amplitude with only two legs being deformed
\begin{align}
 (2\pi)^4\delta^4(\sum_i\hat{p}_i)i\hat{A}_n(z)=(i\lim_{\hat{p}_1^2\rightarrow0}
 \hat{p}_1^2)(i\lim_{\hat{p}_n^2\rightarrow0}\hat{p}_n^2)\hat{G}_2
  \prod_{j=2}^{N-1}(i\lim_{p_j^2\rightarrow0}p_j^2)G_{n-2}. \label{con:LSZ}
\end{align}
Since only $\hat{G}_2$ depends on $z$, then the question of
calculating the $z$-dependence of a deformed amplitude
$\widetilde{B}_n(z)$ is transformed into the calculation of
two-point correlation function of hard fields $\hat{G}_2(z)$ in the
soft background.

The above consideration is limited to the case with only two
external  momenta deformed, and only the terms quadratic in
$\Phi^{\Lambda}$ contribute.  We want to generalize this method to
the multi-leg deformed case, for example, the Risagger deformation
in \cite{Risager:2005vk}. Similar to the two-leg deformed case, the
central part is  corresponding $m$-leg deformed correlation function
\begin{equation}
 \hat{G}_{m}(z)=\int \mathcal{D}\Phi^{\Lambda} \text{exp}\{iS^{\Lambda}[\Phi^{\Lambda},\Phi]\} \Phi_1^{\Lambda}\Phi_2^{\Lambda}\cdots\Phi_m^{\Lambda}.
\end{equation}
where $S^{\Lambda}[\Phi^{\Lambda},\Phi]$ is the hard part of the
expansion of the action after splitting the field. After doing the
LSZ reduction, it will become the large $z$-dependent part of a
$m$-leg deformed amplitude $\widetilde{B}_m(z)$, just like
$\eqref{con:LSZ}$. Particularly, this large $z$-dependent part
$\hat{G}_{m}(z)$ can be calculated by Feynman diagrams of hard
fields in the soft background.

Now we show how to read out Feynman rules in the nontrivial soft
background. After splitting fields into hard part and soft part, we
expand the Lagrangian
\begin{align}
 \mathcal{L}(\tilde{\Phi},\partial_{\mu}\tilde{\Phi})
&=\mathcal{L}(\Phi+\Phi^{\Lambda},\partial_{\mu}\Phi+\partial_{\mu}\Phi^{\Lambda})  \notag\\
&=\mathcal{L}(\Phi,\partial_{\mu}\Phi)+\sum_i\frac{\partial\mathcal{L}}{\partial\tilde{\Phi}_i}\Phi_i^{\Lambda}
    +\sum_i\frac{\partial\mathcal{L}}{\partial(\partial_{\mu}\tilde{\Phi}_i)}\partial_{\mu}\Phi_i^{\Lambda}  \notag\\
&+\frac{1}{2!}\sum_{i,j}\frac{\partial^2\mathcal{L}}{\partial\tilde{\Phi}_i\partial\tilde{\Phi}_j}\Phi_i^{\Lambda}\Phi_j^{\Lambda}
    +\frac{1}{2!}\sum_{i,j}\frac{\partial^2\Phi}{\partial(\partial_{\mu}\tilde{\Phi}_i)\partial(\partial_{\nu}\tilde{\Phi}_j)} \partial_{\mu}\Phi^{\Lambda}_i\partial_{\nu}\Phi_j^{\Lambda} \notag\\
&+\sum_{i,j}\frac{\partial^2\Phi}{\partial\tilde{\Phi}_i\partial(\partial_{\mu}\tilde{\Phi}_j)}\Phi_i^{\Lambda}
    \partial_{\mu}\Phi_j^{\Lambda}+\cdots   \notag \\
&=\mathcal{L}(\Phi,\partial_{\mu}\Phi) \notag \\
&+\frac{1}{2!}\sum_{i,j}\frac{\partial^2\mathcal{L}}{\partial\tilde{\Phi}_i\partial\tilde{\Phi}_j}\Phi_i^{\Lambda}\Phi_j^{\Lambda}
    +\frac{1}{2!}\sum_{i,j}\frac{\partial^2\Phi}{\partial(\partial_{\mu}\tilde{\Phi}_i)\partial(\partial_{\nu}\tilde{\Phi}_j)} \partial_{\mu}\Phi^{\Lambda}_i\partial_{\nu}\Phi_j^{\Lambda}
    +\sum_{i,j}\frac{\partial^2\Phi}{\partial\tilde{\Phi}_i\partial(\partial_{\mu}\tilde{\Phi}_j)}\Phi_i^{\Lambda}\partial_{\mu}\Phi_j^{\Lambda}  \notag\\
&+\frac{1}{3!}\sum_{i,j,k}\frac{\partial^3\mathcal{L}}{\partial\tilde{\Phi}_i\partial\tilde{\Phi}_j\partial\tilde{\Phi}_k}\Phi_i^{\Lambda}\Phi_j^{\Lambda}\Phi_k^{\Lambda}
     +\frac{1}{2!}\sum_{i,j,k}\frac{\partial^3\mathcal{L}}{\partial\tilde{\Phi}_i\partial\tilde{\Phi}_j\partial(\partial_{\mu}\tilde{\Phi}_k)}\Phi_i^{\Lambda}\Phi_j^{\Lambda}\partial_{\mu}\Phi_k^{\Lambda}
     +\cdots,
\end{align}
where $\cdots$ represents higher order terms of $\Phi_i^{\Lambda}$
and we  have used the integration by part and equation of motion in
the third equation to eliminate those terms linear in
$\Phi_i^{\Lambda}$. From the above expansion, we can easily see that
we don't need to consider $\mathcal{L}(\Phi,\partial_{\mu}\Phi)$,
which isn't involved with the hard fields $\Phi_i^{\Lambda}$.

There exists the terms that are quadratic of $\Phi_i^{\Lambda}$,
which  will produce the nontrivial propagators by their inverse,
just like in the ordinary Lagrangian. One non-trivial thing is that
in general all fields $\Phi_i^{\Lambda}$ are mixed together through
the coefficients of quadratic terms, then physically these
coefficients act as different propagators linking with different
fields (or a particle of one type changes into another type in the
process of propagation). And the derivatives in those coefficients
are important for the $z$-dependence of the deformed amplitude,
since the derivatives produce deformed momenta in Feynman rules.

There are also higher order terms of $\Phi_i^{\Lambda}$, which
produce  interaction vertices. When we consider $m$-leg deformed
amplitudes, correspondingly we should consider $m$-leg Feynman
diagrams of hard fields in the soft background. It means that if we
want to compute the $m$-deformed correlation function
$\hat{G}_m(z)$, we should consider the Feynman diagrams with $m$
hard external lines constructed by $k$-point vertex with $k\le m$.
For example, in the $2$-deformed case, we only consider
$S_2^{\Lambda}[\Phi^{\Lambda},\Phi]$ which only gives a propagator
with two hard external line, and in the $3$-deformed case, we should
consider the terms in the expansion of action up to cubic order of
$\Phi_i^{\Lambda}$ and the correspondingly Feynman diagrams is
constructed by three-vertices. In the rest of paper, we will mainly
focus on three-leg deformed case to demonstrate our main idea, and
there are no differences for more general $m$-leg deformed case.

\section{Real Scalar Field Theory} \label{real}
In this section, we will focus on the simplest theory, i.e., the
real scalar field theory, as an example to show our method. More
complicated theories will be considered  in later sections. From now
on, we are limited in three-leg deformed case. The Lagrangian of
real scalar field theory we are considering is
\begin{equation}
  \mathcal{L}=-\frac{1}{2}\partial_{\mu}\phi\partial^{\mu}\phi+\frac{\lambda_{m}}{m!}\phi^{m}, \label{scalar-lagrangian}
\end{equation}
with $m\ge3$.
We make the substitution $\phi\rightarrow\phi+\Phi$ with $\Phi$ representing the hard part of the field and $\phi$ representing the soft part, then the expansion is
\begin{align}
  \mathcal{L} =&-\frac{1}{2}\partial_{\mu}(\phi+\Phi)\partial^{\mu}(\phi+\Phi)+\frac{\lambda_{m}}{m!}(\phi+\Phi)^{m} \notag \\
              =&-\frac{1}{2}\partial_{\mu}\phi\partial^{\mu}\phi+\frac{\lambda_{m}}{m!}\phi^{m} \notag \\
               &-\frac{1}{2}\partial_{\mu}\Phi\partial^{\mu}\Phi+\frac{\lambda_{m}}{2!(m-2)!}\phi^{m-2}\Phi^2+\cdots+\frac{\lambda_{m}}{m!}\Phi^{m} \notag \\
              =&\mathcal{L}(\phi)+\mathcal{L}(\phi,\Phi).
\end{align}
Here we have omitted these terms proportional to $\Phi$ following
the  same arguments in previous section. $\mathcal{L}(\phi)$
represents the soft part of Lagrangian, and $\mathcal{L}(\phi,\Phi)$
represents the hard part of Lagrangian, called hard Lagrangian.

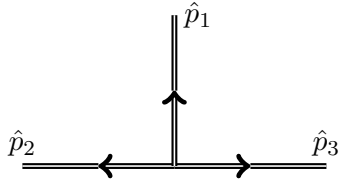
\begin{figure}
  \begin{center}
    \begin{tikzpicture}
      \draw [double,thick,->] (0,0)--(1,0);
      \draw [double,thick] (1,0)--(2,0);
      \draw [double,thick] (-2,0)--(-1,0);
      \draw [double,thick,<-] (-1,0)--(0,0);
      \draw [double,thick,->] (0,0)--(0,1);
      \draw [double,thick] (0,1)--(0,2);
      \node [right] at (0,2) {$\hat{p}_1$};
      \node [above] at (-2,0) {$\hat{p}_2$};
      \node [above] at (2,0) {$\hat{p}_3$};
    \end{tikzpicture}
  \end{center}
  \caption{The Feynman digram in soft background contributing for $\langle\Phi_1\Phi_2\Phi_3\rangle$. Here double solid lines represent the propagators of hard particles in soft background.}  \label{con:scalarcase}
\end{figure}

We regroup the $\mathcal{L}(\phi,\Phi)$ into the quadratic part and
the higher  order part, then the quadratic terms of $\Phi$ produce
the propagator and the other terms give the interactive vertices
\begin{align}
  \mathcal{L}(\phi,\Phi)=&\frac{1}{2}\Phi\displaystyle\biggl(\partial^2+\frac{\lambda_{m}}{(m-2)!}\phi^{m-2}\displaystyle\biggl)\Phi+\frac{\lambda_{m}}{3!(m-3)!}\phi^{m-3}\Phi^3+\cdots+\frac{\lambda_{m}}{m!}\Phi^{m}\notag\\
                        =& \frac{1}{2}\Phi D\Phi+\mathcal{L}_1(\phi,\Phi).
\end{align}
Here we have used integration by parts and ignored the total
derivative. We  can read out the Feynman rules for propagator and
vertices from the hard Lagrangian directly: the propagator is
$-iD^{-1}$ with $\partial^2$ replaced by $-p^2$ and the three-vertex
is $i\frac{\lambda_{m}}{(m-3)!}\phi^{m-3}$. Because we only consider
three-leg deformation, then only cubic vertex can contribute. And
for the deformed correlation function,
$\langle\Phi_1\Phi_2\Phi_3\rangle$, there is only one contributing
Feynman diagram with a cubic vertex and three hard propagator lines
as shown in Figure \ref{con:scalarcase}.

\begin{figure}[h]
\centering
\includegraphics[scale=0.3]{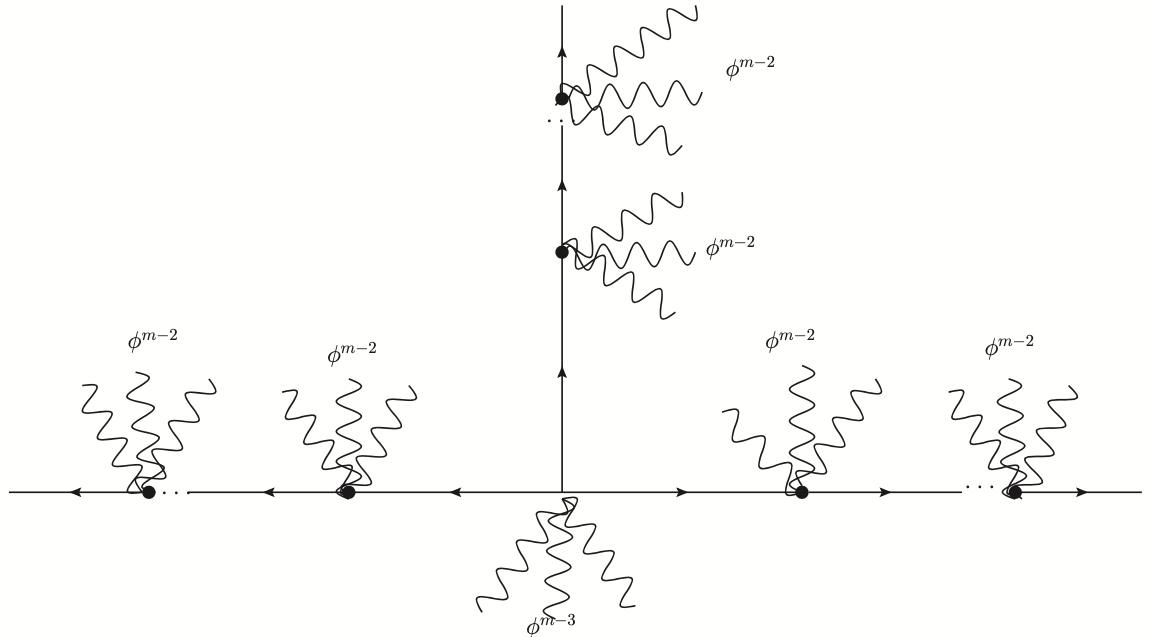}
\caption{The Feynman diagram represents the real scatting process without soft background after expanding the double-line propagators in Figure \ref{con:scalarcase}.} \label{con:expansiondiagram}
\end{figure}

Now we calculate explicitly the $z$-dependent part of a $n$-point
amplitude of  real scalar theory. After doing the LSZ reduction, the
deformed correlation function $\hat{G}_3(z)$ becomes
$\widetilde{B}_3(z)$
\begin{align}
  i\widetilde{B}_3(z)=&(i\lim_{\hat{p}_1^2\rightarrow0}\hat{p}_1^2)(i\lim_{\hat{p}_2^2\rightarrow0}\hat{p}_2^2)(i\lim_{\hat{p}_3^2\rightarrow0}\hat{p}_3^2)\hat{G}_3(z)\notag \\
        =& i\lim_{\hat{p}_1^2\rightarrow0}\lim_{\hat{p}_2^2\rightarrow0}\lim_{\hat{p}_3^2\rightarrow0} \hat{p}_1^2\hat{p}^2_2\hat{p}^2_3 D^{-1}_1D^{-1}_2D^{-1}_3 \frac{\lambda_{m}}{(m-3)!}\phi^{m-3} \notag\\
        =& i  \frac{1}{-1+\frac{1}{\hat{P}_1^2}\frac{\lambda_{m}}{(m-2)!}\phi^{m-2}} \frac{1}{-1+\frac{1}{\hat{P}_2^2}\frac{\lambda_{m}}{(m-2)!}\phi^{m-2}} \frac{1}{-1+\frac{1}{\hat{P}_3^2}\frac{\lambda_{m}}{(m-2)!}\phi^{m-2}} \frac{\lambda_{m}}{(m-3)!}\phi^{m-3}\notag\\
        =& -i\frac{\lambda_{m}\phi^{m-3}}{(m-3)!} \sum_{i=0}^{\infty} \sum_{j=0}^{\infty}\sum_{k=0}^{\infty} \big[\frac{\lambda_{m}}{(m-2)!}\phi^{m-2}\frac{1}{\hat{P}_1^2}\big]^i \big[\frac{\lambda_{m}}{(m-2)!}\phi^{m-2}\frac{1}{\hat{P}_2^2}\big]^j  \big[\frac{\lambda_{m}}{(m-2)!}\phi^{m-2}\frac{1}{\hat{P}_3^2}\big]^k\notag\\
        =&-i\frac{\lambda_{m}\phi^{m-3}}{(m-3)!} \big[ 1+\frac{\lambda_{m}}{(m-2)!}\phi^{m-2}\frac{1}{\hat{P}_1^2}+\frac{\lambda_{m}}{(m-2)!}\phi^{m-2}\frac{1}{\hat{P}_2^2}+\frac{\lambda_{m}}{(m-2)!}\phi^{m-2}\frac{1}{\hat{P}_3^2}+\cdots \big] \label{con:v3expansion}
\end{align}
where we expanded the propagators and dots represents higher order
terms in the expansion. Now we interpret the physical meaning of the
above formula.  In \eref{con:v3expansion}, each double-line
propagator in soft background is expanded by geometric series with
each term being a product of a propagator $1/\hat{P}^2$ and a vertex
$\phi^{m-2}$ to some power, and the expansion can be depicted by the
diagram in Figure \ref{con:expansiondiagram}. In the diagram
\ref{con:expansiondiagram}, every solid line represents a propagator
of the hard field without soft background, every two-vertex
represents a real vertex connected with $(m-2)$ soft lines and $2$
hard lines and the cubic vertex connected with $(m-3)$ soft lines
and $3$ hard lines because of the interaction term
$\frac{\lambda_{m}}{3!(m-3)!}\phi^{m-3}\Phi^3$. Then we can see that
the diagram in Figure \ref{con:scalarcase} exactly describes the
propagation and interaction of hard particles in the soft
background. So $\widetilde{B}_3(z)$ represents the complete
contribution of all Feynman diagrams containing the boundary part of
deformed amplitude $\hat{A}_n(z)$.

From the general picture, we can specify the boundary behavior or
large $z$ behavior of deformed amplitude $\hat{A}_n(z)$. If we
choose the three deformed momenta $\hat{p}_i(z)=p_i+zq_i$ with
$i=1,2,3$ satisfying the conditions
\begin{align}
      q_i\cdot q_j=0~ \text{with}~ i,j=1,2,3; \qquad p_i\cdot q_i=0, ~\text{with}~ i=1,2,3,
\end{align}
we will get $\hat{p}_i^2=0$, i.e., the on-shell condition for three
hard legs.  We should emphasize that although $\hat{p}_i^2=0$, the
expansion in \eref{con:v3expansion} is $\hat{P}_i$, which contains
not only the hard momentum $\hat{p}_i$, but some soft momenta
$p_j$'s ($j=4,\cdots,n$) for the vertex
$\frac{\lambda_{m}}{(m-2)!}\phi^{m-2}$. With this explanation, we
see that
\begin{align}
     \frac{1}{(\hat{p}_i+P_S)^2}=\frac{1}{(p_i+zq_i+P_S)^2}=\frac{1}{2P_S\cdot p_i+2zP_S\cdot q_i+P_S^2 }  \sim \frac{1}{z},
\end{align}
where $P_S=\sum_{j\in S}p_j$ with $S\subset \{4,\cdots,n\}$ and in
the  last step we take $z\rightarrow \infty$. Then the contribution
of every propagator $1/\hat{P}_i^2$ in \eref{con:v3expansion} is
$O(\frac{1}{z})$, so we can easily see that the leading term of the
expansion, $\frac{\lambda_{m}}{(m-3)!}\phi^{m-3}$, is in the order
$O(1)$ and all other terms in the expansion vanish for the
additional propagators $1/\hat{P}_i^2$. We conclude that
$\widetilde{B}_3(z)$ is non-vanishing in the three-leg deformed
amplitude of real scalar theory.

Although the above method is general, however the three-leg deformed
case  is a little special since the diagram in Figure
\ref{con:scalarcase} has only a cubic vertex and doesn't have any
internal propagators. Then a question arises: what happens if we
make a four-deformation or even higher $s$-deformation? Now let's
assume we have made a $s$-leg deformation with $s\ge 4$. The
situation becomes a little different for there are two cases $m\ge
s$ or $m<s$, where $m$ is the power of the leading term in
\eref{scalar-lagrangian}. If $m\ge s$, then in the expansion of
Lagrangian there is always a term $\phi^{m}\Phi^{m-s}$ which will
give a Feynman diagram having only one $(m-s)$-leg vertex, and the
boundary contribution of this diagram doesn't vanish, so in this
case the boundary contribution doesn't vanish for the $s$-leg
deformed on-shell recursion relation. For the second case $m<s$, all
interaction terms in the expansion of Lagrangian are like
$\phi^{m}\Phi^{m-t}$ with $t<s$, so all contributing diagrams for
the $z$-dependent part of $s$-leg deformed amplitude must have a
extra hard propagator which act as $O(\frac{1}{z})$, then in this
case the boundary contribution vanishes. So for the real scalar
Lagrangian with finite terms, we can always use on-shell recursion
relations with enough many legs being deformed, whose boundary term
vanishes.\footnote{Using the terminology in other literatures, the
real scalar theory with finite interaction terms is always on-shell
constructible.}

%%%%%%%%%%%%%%%%%%%
\section{Yukawa Theory} \label{yukawa}
%%%%%%%%%%%%%%%%%%%%

In this section, we will move on to consider a little more
complicated theory,  i.e. the Yukawa theory. The Lagrangian of the
Yukawa theory considered here is\footnote{Here we don't consider the
interaction of saclar field.}
\begin{align}
  \mathcal{L}=-\frac{1}{2}\partial^{\mu}\phi\partial_{\mu}\phi+i\bar\psi\gamma^{\mu}\partial_{\mu}\psi+\lambda\bar\psi\psi\phi.
\end{align}
In comparison with the real scalar field theory, the Yukawa theory
have  some differences. First, the appearance of fermionic fields
$\psi$ and $\bar\psi$ bring some extra minus signs in the process of
calculation for commutating two fermionic fields. Second, since the
number of fields is more than one and they interact with each other,
then after doing the expansion, there will appear some propagators
connecting different fields.\footnote{It is because we are only
considering the hard fields and ignoring the soft fields, there
appear such unphysical propagators. But if we do the series
expansion of these propagators, then we will find they are actually
physical, just like the Figure $\ref{con:expansiondiagram}$.}
Because of these reasons, the number of diagrams we should consider
will be more than one.

Just like before, we split the fields into hard parts and soft
parts:  $\phi\rightarrow H+\phi, \bar{\psi}\rightarrow
\bar{\Psi}+\psi,\psi\rightarrow \Psi+\psi$, then the Lagrangian are
also divided into two parts:
\begin{align}
  \mathcal{L}=&\mathcal{L}(\phi,\psi,\bar\psi)+\mathcal{L}(H,\bar\Psi,\Psi) \notag \\
             =&-\frac{1}{2}\partial^{\mu}\phi\partial_{\mu}\phi+i\bar\psi\gamma^{\mu}\partial_{\mu}\psi+\lambda\bar\psi\psi\phi \notag\\
              &-\frac{1}{2}\partial^{\mu}H\partial_{\mu}H+i\bar\Psi\gamma^{\mu}\partial_{\mu}\Psi+\lambda\bar\psi\Psi H+\lambda\bar\Psi\psi H
               +\lambda\bar\Psi\Psi\phi+\lambda\bar\Psi\Psi H,
\end{align}
where we have used equations of motion to drop those terms proportional to hard fields.
Then we only focus on the hard part of Lagrangian and recast it as,
\begin{align}
  \mathcal{L}(H,\bar\Psi,\Psi)
     =&-\frac{1}{2}\partial^{\mu}H\partial_{\mu}H+i\bar\Psi\gamma^{\mu}\partial_{\mu}\Psi+\lambda\bar\psi\Psi H+\lambda\bar\Psi\psi H+\lambda\bar\Psi\Psi\phi+\lambda\bar\Psi\Psi H \notag \\
     =&-\frac{1}{2} \left( \begin{array}{ccc} H &\bar\Psi &\Psi^T \end{array} \right)
            \left( \begin{array}{ccc} -\partial^2 &\lambda\psi^T &-\lambda\bar\psi \\
                      -\lambda\psi &0 &-i\gamma^{\mu}\overrightarrow{\partial}_{\mu}-\lambda\phi \\
                     \lambda\bar\psi^{T} &i(\gamma^{\mu})^T\overleftarrow{\partial}_{\mu}+\lambda\phi &0 \end{array} \right)
            \left( \begin{array}{c} H\\ \bar\Psi^{T}\\ \Psi \end{array} \right)
    +\lambda\bar\Psi\Psi H \notag\\
    =&-\frac{1}{2} \left( \begin{array}{ccc} H &\bar\Psi &\Psi^T \end{array} \right) D \left( \begin{array}{c} H\\ \bar\Psi^{T}\\ \Psi \end{array} \right)
        +\lambda\bar\Psi\Psi H  \notag\\
    =&\mathcal{L}_0(H,\Psi,\bar\Psi)+\mathcal{L}_1(H,\Psi,\bar\Psi),
\end{align}
where we have used integration by parts and transposition of
fermionic fields  like $\bar{\Psi}\Psi=-\Psi^{T}\bar{\Psi}^T$,
$\bar{\Psi}\gamma^{\mu}\partial_{\mu}\Psi=-(\partial_{\mu}\Psi^T)(\gamma^{\mu})^T\bar{\Psi}^T$.
Here $\mathcal{L}_0(H,\Psi,\bar\Psi)$ represents the free part of
hard Lagrangian, from which we can get the expressions of
propagators, and $\mathcal{L}_1(H,\Psi,\bar\Psi)$ is the interaction
part, which gives use the cubic vertex. And we can simply infer that
there are six propagators for $D$ is antisymmetric.

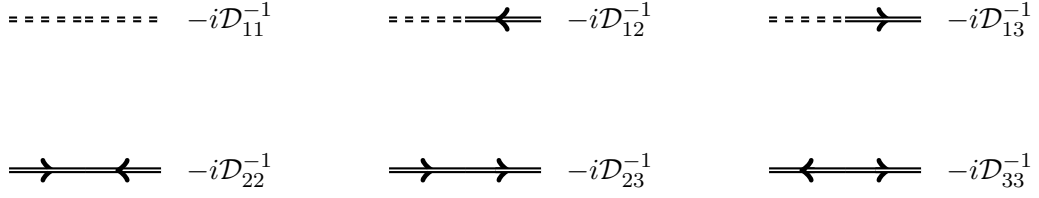
\begin{figure}
  \begin{center}
  \begin{tikzpicture}
  %%%%%%%% first line
        \draw [double,dashed,thick] (0,0)--(1,0);
        \draw [double,dashed,thick] (1,0)--(2,0);
        \node [right] at (2.2,0) {$-i\mathcal{D}_{11}^{-1}$};

        \draw [double,thick, dashed] (5,0)--(6,0);
        \draw [double,thick] (6,0)--(6.5,0);
        \draw [double,thick,<-] (6.4,0)--(7,0);
        \node [right] at (7.2,0) {$-i\mathcal{D}^{-1}_{12}$};

        \draw [double,thick, dashed] (10,0)--(11,0);
        \draw [double,thick,->] (11,0)--(11.6,0);
        \draw [double,thick] (11.5,0)--(12,0);
        \node [right] at (12.2,0) {$-i\mathcal{D}^{-1}_{13}$};
  %%%%%% second line
        \draw [double,thick,->] (0,-2)--(0.6,-2);
        \draw [double,thick] (0.5,-2)--(1,-2);
        \draw [double,thick] (1,-2)--(1.5,-2);
        \draw [double,thick,<-] (1.4,-2)--(2,-2);
        \node [right] at (2.2,-2) {$-i\mathcal{D}_{22}^{-1}$};

        \draw [double,thick,->] (5,-2)--(5.6,-2);
        \draw [double,thick] (5.5,-2)--(6,-2);
        \draw [double,thick,->] (6,-2)--(6.6,-2);
        \draw [double,thick] (6.4,-2)--(7,-2);
        \node [right] at (7.2,-2) {$-i\mathcal{D}_{23}^{-1}$};

        \draw [double,thick] (10,-2)--(10.6,-2);
        \draw [double,thick,<-] (10.4,-2)--(11,-2);
        \draw [double,thick,->] (11,-2)--(11.6,-2);
        \draw [double,thick] (11.4,-2)--(12,-2);
        \node [right] at (12.2,-2) {$-i\mathcal{D}_{33}^{-1}$};
  \end{tikzpicture}
  \end{center}
  \caption{The six propagators for hard fields in soft background. Here a dashed double-line represents a propagating scalar boson, while a solid double-line represents a propagating fermion, and the arrow represents the direction of the propagation of a fermion.} \label{con:Yukawap}
  \end{figure}
To get the concrete expressions for these propagators,
we first divide the matrix $D$ into two parts $D_0$ and $V$ because the inverse of $D_0$ is easy to calculate, then we apply the geometric series expansion to $D$
\begin{align}
  D^{-1}=&(D_0+V)^{-1} \notag\\
        =&[D_0(1+D_0^{-1}V)]^{-1} \notag\\
        =&\sum_{k=0}^{\infty}(-1)^k(D_0^{-1}V)^kD_0^{-1} \notag \\
        =&D_0^{-1}-D_0^{-1}VD_0^{-1}+D_0^{-1}VD_0^{-1}VD_0^{-1}+\cdots, \label{expansion-D}
\end{align}
with
\begin{align}
  D_0=\left( \begin{array}{ccc} -\partial^2 &0 &0 \\
                                     0 &0 &-i \overrightarrow{\slashed{\partial}}\\
                                     0 &i\overleftarrow{\slashed{\partial}}^T &0 \end{array} \right),
\quad
 V=\left( \begin{array}{ccc}  0&\lambda\psi^T &-\lambda\bar\psi \\
                                   -\lambda\psi &0 &-\lambda\phi \\
                                   \lambda\bar\psi^{T} &\lambda\phi &0 \end{array} \right),
\end{align}
where we have used the formula of inverse of a multiplication of two
operators, $(AB)^{-1}=B^{-1}A^{-1}$. The inverse of $D_0$ is easy to
get as
\begin{align}
  D_0^{-1}=\left( \begin{array}{ccc} -\frac{1}{\partial^2} &0 &0 \\
                                     0 &0 &\frac{-i\overleftarrow{\slashed{\partial}}^T}{\partial^2}\\
                                     0 &\frac{i\overrightarrow{\slashed{\partial}}}{\partial^2} &0 \end{array} \right),
\end{align}
where the arrows represent the directions of the action of the derivatives in numerators,
and we can easily check $D_0 D_0^{-1}=D_0^{-1}D_0=1$.\footnote{We choose the convention $\{\gamma^{\mu},\gamma^{\nu}\}=2g^{\mu\nu}$.}
So the inverse of $D$ is given by the expansion \eref{expansion-D} as
\begin{align}
&D^{-1}=\left( \begin{array}{ccc} -\frac{1}{\partial^2} &0 &0 \\
  0 &0 &\frac{-i\overleftarrow{\slashed{\partial}}^T}{\partial^2}\\
  0 &\frac{i\overrightarrow{\slashed{\partial}}}{\partial^2} &0 \end{array} \right)
-\left( \begin{array}{ccc} 0
  &-\frac{1}{\partial^2}(-\lambda\bar\psi)\frac{i\overrightarrow{\slashed{\partial}}}{\partial^2}
  &-\frac{1}{\partial^2}(\lambda\psi^T)\frac{-i\overleftarrow{\slashed{\partial}}^T}{\partial^2} \\
  \frac{-i\overleftarrow{\slashed{\partial}}^T}{\partial^2}(\lambda\bar\psi^T)(-\frac{1}{\partial^2})
  &0
  &\frac{-i\overleftarrow{\slashed{\partial}}^T}{\partial^2}(\lambda\phi)\frac{-i\overleftarrow{\slashed{\partial}}^T}{\partial^2}\\
  \frac{i\overrightarrow{\slashed{\partial}}}{\partial^2}(-\lambda\psi)(-\frac{1}{\partial^2})
  &\frac{i\overrightarrow{\slashed{\partial}}}{\partial^2}(-\lambda\phi)\frac{i\overrightarrow{\slashed{\partial}}}{\partial^2}
  &0 \end{array} \right) + \notag\\
&\left( \begin{array}{cc} -\frac{1}{\partial^2}(-\lambda\bar\psi)\frac{i\overrightarrow{\slashed{\partial}}}{\partial^2}(-\lambda\psi)(-\frac{1}{\partial^2})
    -\frac{1}{\partial^2}(\lambda\psi^T)\frac{-i\overleftarrow{\slashed{\partial}}^T}{\partial^2}(\lambda\bar\psi^T)(-\frac{1}{\partial^2})
   &-\frac{1}{\partial^2}(-\lambda\bar\psi)\frac{i\overrightarrow{\slashed{\partial}}}{\partial^2}(-\lambda\phi)\frac{i\overrightarrow{\slashed{\partial}}}{\partial^2}
    \\
   \frac{-i\overleftarrow{\slashed{\partial}}^T}{\partial^2}(\lambda\phi)\frac{-i\overleftarrow{\slashed{\partial}}^T}{\partial^2}(\lambda\bar\psi^T)(-\frac{1}{\partial^2})
   &\frac{-i\overleftarrow{\slashed{\partial}}^T}{\partial^2}(\lambda\bar\psi^T)(-\frac{1}{\partial^2})(-\lambda\bar\psi)\frac{i\overrightarrow{\slashed{\partial}}}{\partial^2}
  \\
   \frac{i\overrightarrow{\slashed{\partial}}}{\partial^2}(-\lambda\phi)\frac{i\overrightarrow{\slashed{\partial}}}{\partial^2}(-\lambda\psi)(-\frac{1}{\partial^2})
   &\frac{i\overrightarrow{\slashed{\partial}}}{\partial^2}(-\lambda\psi)(-\frac{1}{\partial^2})(-\lambda\bar\psi)\frac{i\overrightarrow{\slashed{\partial}}}{\partial^2}
   +\frac{i\overrightarrow{\slashed{\partial}}}{\partial^2}(-\lambda\phi)\frac{i\overrightarrow{\slashed{\partial}}}{\partial^2}(-\lambda\phi)\frac{i\overrightarrow{\slashed{\partial}}}{\partial^2}
 \end{array} \right. \notag\\
 &  \left. \begin{array}{cc}
 &-\frac{1}{\partial^2}(\lambda\psi^T)\frac{-i\overleftarrow{\slashed{\partial}}^T}{\partial^2}(\lambda\phi)\frac{-i\overleftarrow{\slashed{\partial}}^T}{\partial^2}\\
 &\frac{-i\overleftarrow{\slashed{\partial}}^T}{\partial^2}(\lambda\bar\psi^T)(-\frac{1}{\partial^2})(\lambda\psi^T)\frac{-i\overleftarrow{\slashed{\partial}}^T}{\partial^2}
  +\frac{-i\overleftarrow{\slashed{\partial}}^T}{\partial^2}(\lambda\phi)\frac{-i\overleftarrow{\slashed{\partial}}^T}{\partial^2}(\lambda\phi)\frac{-i\overleftarrow{\slashed{\partial}}^T}{\partial^2}\\
 &\frac{i\overrightarrow{\slashed{\partial}}}{\partial^2}(-\lambda\psi)(-\frac{1}{\partial^2})(\lambda\psi^T)\frac{-i\overleftarrow{\slashed{\partial}}^T}{\partial^2}\\
\end{array} \right)
   +\cdots, \label{con:D}
\end{align}
where $\cdots$ represents the higher order terms in the expansion \eref{expansion-D}. From the above formula \eref{con:D}, we should note that there is totally $n$ derivatives $-\frac{1}{\partial^2}$, $\frac{-i\overleftarrow{\slashed{\partial}}^T}{\partial^2}$ or $\frac{i\overrightarrow{\slashed{\partial}}}{\partial^2}$ multiplied together in every element of the $n$th order matrix, and the frst order has no $\lambda$, the second order elements are linear of $\lambda$, then elements of $n$th order matrix should be proportional to $\lambda^{n-1}$.\footnote{The two facts are consistent with \eref{expansion-D}, since every element of $V$ has a $\lambda$ and derivatives are only contained by $D_0^{-1}$.}  And we should note that $D^{-1}_{22},D^{-1}_{33}\ne 0$ for the corrections of high order terms which is different from the ordinary Yukawa theory.

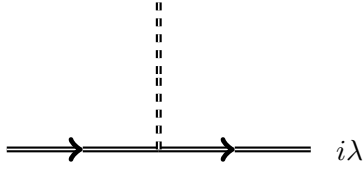
\begin{figure}
  \begin{center}
    \begin{tikzpicture}
      \draw [double,thick,->] (0,0)--(1,0);
      \draw [double,thick] (1,0)--(2,0);
      \draw [double,thick,->] (-2,0)--(-1,0);
      \draw [double,thick] (-1,0)--(0,0);
      \draw [double,thick,dashed] (0,0)--(0,1);
      \draw [double,thick,dashed] (0,1)--(0,2);
      \node [right] at (2.2,0) {$i\lambda$};
    \end{tikzpicture}
  \end{center}
  \caption{Three-vertex in soft background for Yukawa theory, $i\lambda$.}  \label{con:Yukawav}
\end{figure}

After replacing $\slashed{\partial}, \partial^2$ by $i\slashed{P},
-P^2$, then we can get the concrete expressions of all propagators
from  $\eqref{con:D}$, as shown in Figure \ref{con:Yukawap}, where
$\mathcal{D}_{ij}=D_{ij}^{-1}(\partial\rightarrow P)$. We can also
derive the expression of vertex of the hard Lagrangian, which are
drawn in Figure \ref{con:Yukawav} following the conventions made
above.
Just as in the previous seciton, we choose to deform the same momenta and impose the same conditions, then the large $z$ behaviors of elements of $D_0^{-1}$ are
\begin{align}
 -\frac{1}{\partial^2}\sim \frac{1}{\hat{P}^2}=\frac{1}{P^2+2P\cdot q}\sim \frac{1}{z},
 \quad
 \frac{\slashed{\partial}}{\partial^2}\sim \frac{\hat{\slashed{P}}}{\hat{P}^2}=
 \frac{\slashed{P}+z\slashed{q}}{P^2+2zq\cdot P}\sim \frac{\slashed{q}}{2q\cdot P},
\end{align}
since the elements of $V$ don't contribute, so deformed correlation
function  is only dependant on $D_0^{-1}$ in the expansion of
$D^{-1}$. When we consider higher order terms in the expansion of
$D^{-1}$,  new things appear
\begin{align}
  \frac{\slashed{\partial}_1}{\partial_1^2} \frac{\slashed{\partial}_2}
  {\partial_2^2}=\frac{\partial_1\cdot \partial_2}{\partial_1^2 \partial_2^2}\sim
  \frac{\hat{p}_1\cdot \hat{p}_2}{\hat{p}_1^2 \hat{p}_2^2}=\frac{p_1\cdot p_2
  +2z(p_1\cdot q_2+p_2\cdot q_1)}{(p_1^2+2zp_1\cdot q_1)(p_2^2+2zp_2\cdot q_2)}
   \sim \frac{1}{z},~~~\Label{arg-1}
\end{align}
where we have used $q_1\cdot q_2=0$. From the above results, we can
conclude  that non-vanishing elements of the second order
$D_0^{-1}VD_0^{-1}$ behave as $1/z$ when $z\rightarrow \infty$, and
elements of higher order terms vanish even faster. So when we
consider the large $z$ behavior of deformed correlation functions,
we should focus on only one propagators $-i\mathcal{D}_{23}^{-1}$,
since only the first order term of the expansion of the propagator
may contribute.
%While if we consider the boundary behavior of amplitude, because of LSZ reduction, the vanishing $z$ behavior of
% some deformed correlation functions will be suppressed by
% $\hat{p}_i^2\sim z$ partly.
\\

Now we begin to calculate the $z$-dependence of deformed amplitudes
$\widetilde{B}_3(z)$ explicitly. For cases with  three-leg
deformation, there are only four cases $\langle H_1H_2H_3\rangle$,
$\langle H_1\Psi_{2\alpha}\bar\Psi_{3\beta}\rangle$, $\langle
\Psi_{1\alpha}\bar\Psi_{2\beta}\Psi_{3\gamma}\rangle$ and $\langle
\bar\Psi_{1\alpha}\Psi_{2\beta}\bar\Psi_{3\gamma}\rangle$, and the
third one $\langle
\Psi_{1\alpha}\bar\Psi_{2\beta}\Psi_{3\gamma}\rangle$ is just the
Hermitian conjugate of the fourth one $\langle
\bar\Psi_{1\alpha}\Psi_{2\beta}\bar\Psi_{3\gamma}\rangle$. The
procedure is the same as in the previous section: we will draw the
corresponding Feynman diagrams, then write down the expressions, and
calculate $\widetilde{B}_3(z)$ by LSZ reduction.

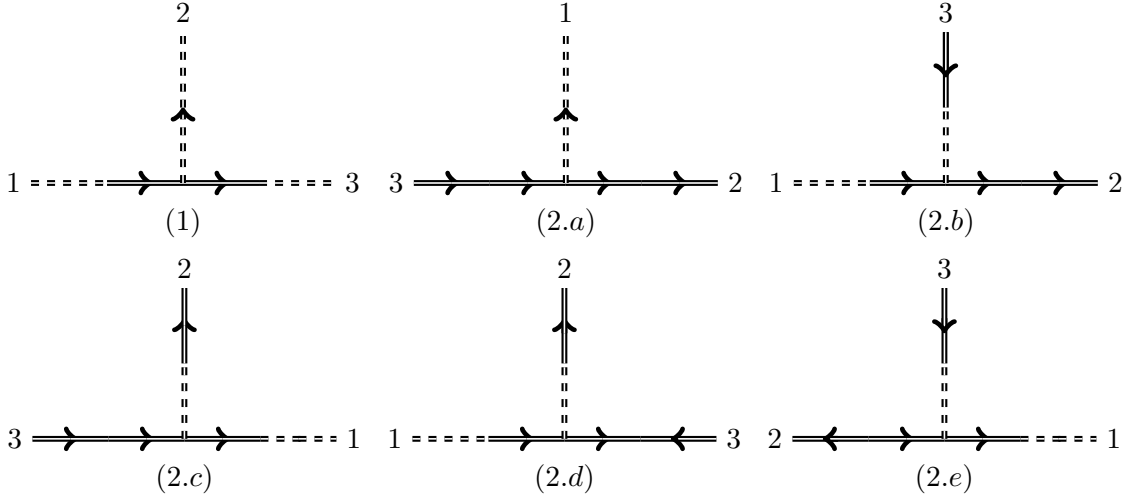
\begin{figure}
  \begin{center}
    \begin{tikzpicture}
      %%%%% diagrams for the first case
      \draw [double,thick,->] (0,0)--(0.6,0);
      \draw [double,thick] (0.5,0)--(1,0);
      \draw [double,dashed,thick] (1,0)--(2,0);
      \draw [double,dashed,thick] (-2,0)--(-1,0);
      \draw [double,thick,->] (-1,0)--(-0.4,0);
      \draw [double,thick] (-0.5,0)--(0,0);
      \draw [double,thick,dashed,->] (0,0)--(0,1);
      \draw [double,thick,dashed] (0,1)--(0,2);
      \node [left] at (-2,0) {$1$};
      \node [above] at (0,2) {$2$};
      \node [right] at (2.0,0) {$3$};
      \node [below] at (0,-0.2) {$(1)$};
    \end{tikzpicture}
    \begin{tikzpicture}
      %%%%%%% diagrams for the second case
      \draw [double,thick,->] (0,0)--(0.6,0);
      \draw [double,thick] (0.5,0)--(1,0);
      \draw [double,thick,->] (1,0)--(1.6,0);
      \draw [double,thick] (1.5,0)--(2,0);
      \draw [double,thick,->] (-2,0)--(-1.4,0);
      \draw [double,thick] (-1.5,0)--(-1,0);
      \draw [double,thick,->] (-1,0)--(-0.4,0);
      \draw [double,thick] (-0.5,0)--(0,0);
      \draw [double,thick,dashed,->] (0,0)--(0,1);
      \draw [double,thick,dashed] (0,1)--(0,2);
      \node [left] at (-2,0) {$3$};
      \node [above] at (0,2) {$1$};
      \node [right] at (2,0) {$2$};
      \node [below] at (0,-0.2) {$(2.a)$};
      %%%%%%
      \draw [double,thick,->] (5,0)--(5.6,0);
      \draw [double,thick] (5.5,0)--(6,0);
      \draw [double,thick,->] (6,0)--(6.6,0);
      \draw [double,thick] (6.5,0)--(7,0);
      \draw [double,dashed,thick] (3,0)--(4,0);
      \draw [double,thick,->] (4,0)--(4.6,0);
      \draw [double,thick] (4.5,0)--(5,0);
      \draw [double,thick,dashed] (5,0)--(5,1);
      \draw [double,thick] (5,1)--(5,2);
      \draw [double,thick,<-] (5,1.4)--(5,2);
      \node [left] at (3,0) {$1$};
      \node [above] at (5,2) {$3$};
      \node [right] at (7,0) {$2$};
      \node [below] at (5,-0.2) {$(2.b)$};
    \end{tikzpicture}

    \begin{tikzpicture}
      %%%%%
      \draw [double,thick,->] (0,0)--(0.6,0);
      \draw [double,thick] (0.5,0)--(1,0);
      \draw [double,thick,dashed] (1,0)--(1.6,0);
      \draw [double,thick,dashed] (1.5,0)--(2,0);
      \draw [double,thick,->] (-2,0)--(-1.4,0);
      \draw [double,thick] (-1.5,0)--(-1,0);
      \draw [double,thick,->] (-1,0)--(-0.4,0);
      \draw [double,thick] (-0.5,0)--(0,0);
      \draw [double,thick,dashed] (0,0)--(0,1);
      \draw [double,thick,->] (0,1)--(0,1.6);
      \draw [double,thick] (0,1.5)--(0,2);
      \node [left] at (-2,0) {$3$};
      \node [above] at (0,2) {$2$};
      \node [right] at (2,0) {$1$};
      \node [below] at (0,-0.2) {$(2.c)$};
       %%%%%%
       \draw [double,thick,->] (5,0)--(5.6,0);
       \draw [double,thick] (5.5,0)--(6,0);
       \draw [double,thick] (6,0)--(6.5,0);
       \draw [double,thick,<-] (6.4,0)--(7,0);
       \draw [double,dashed,thick] (3,0)--(4,0);
       \draw [double,thick,->] (4,0)--(4.6,0);
       \draw [double,thick] (4.5,0)--(5,0);
       \draw [double,thick,dashed] (5,0)--(5,1);
       \draw [double,thick,->] (5,1)--(5,1.6);
       \draw [double,thick] (5,1.4)--(5,2);
       \node [left] at (3,0) {$1$};
       \node [above] at (5,2) {$2$};
       \node [right] at (7,0) {$3$};
       \node [below] at (5,-0.2) {$(2.d)$};
       %%%%%%%
      \draw [double,thick,->] (10,0)--(10.6,0);
      \draw [double,thick] (10.5,0)--(11,0);
      \draw [double,thick,dashed] (11,0)--(11.6,0);
      \draw [double,thick,dashed] (11.5,0)--(12,0);
      \draw [double,thick] (8,0)--(8.6,0);
      \draw [double,thick,<-] (8.4,0)--(9,0);
      \draw [double,thick,->] (9,0)--(9.6,0);
      \draw [double,thick] (9.5,0)--(10,0);
      \draw [double,thick,dashed] (10,0)--(10,1);
      \draw [double,thick] (10,1)--(10,1.6);
      \draw [double,thick,<-] (10,1.4)--(10,2);
      \node [left] at (8,0) {$2$};
      \node [above] at (10,2) {$3$};
      \node [right] at (12,0) {$1$};
      \node [below] at (10,-0.2) {$(2.e)$};
    \end{tikzpicture}
\end{center}
\caption{The Feynman diagram $(1)$ is for $\langle H_1H_2H_3\rangle$ in soft background, while the last digarms are all for $\langle H_1\Psi_{2\alpha}\bar\Psi_{3\beta}\rangle$.} \label{con:Yukawacaseonetwo}
\end{figure}

For the first case $\langle H_1H_2H_3\rangle$, its corresponding
Feynman diagrams are shown in Figure \ref{con:Yukawacaseonetwo},
where we only draw one digram since the other two diagrams can be
got by making permutations of $(123)$. Then the deformed correlation
function is
\begin{align}
  \hat{G}_3(z)=i\lambda (-i\mathcal{D}_{12})(-i\mathcal{D}_{13})(-i\mathcal{D}_{11})+\mathcal{P}(123), \label{con:Yukawafeq}
\end{align}
with $\mathcal{P}(123)$ represent the  terms got by permuting the
three propagators.\footnote{Note that we are not distinguishing the
propagators in position or momentum space, but it doesn't matter
because we will always apply LSZ reduciton and only use the
propagators in momentum space.} Under the LSZ reduction, the
deformed correlation function gives the $z$-dependence of deformed
amplitude as
\begin{align}
i\widetilde{B}_3(z)=&(i\lim_{\hat{p}_1^2\rightarrow0}\hat{p}_1^2)(i\lim_{\hat{p}_2^2\rightarrow0}\hat{p}_2^2)(i\lim_{\hat{p}_3^2\rightarrow0}\hat{p}_3^2)\hat{G}_3(z)+\mathcal{P}(123)\notag \\
=& i\lambda \lim_{\hat{p}_1^2\rightarrow0} \lim_{\hat{p}_2^2\rightarrow0}
\lim_{\hat{p}_3^2\rightarrow0}\hat{p}_1^2\hat{p}_2^2\hat{p}_3^2
 [\frac{1}{\hat{P}_3^2}(\lambda\bar{\psi})\frac{-\slashed{\hat{P}}_3}{\hat{P}_3^2}+\cdots]
 [\frac{1}{\hat{P}_2^2}+\cdots] [\frac{1}{\hat{P}_1^2}(\lambda\psi^T)\frac{-\slashed{\hat{P}}_1}{\hat{P}_1^2}+\cdots]+\mathcal{P}(123) \notag\\
 =& i\lambda
 [(\lambda\bar{\psi})\frac{-\slashed{\hat{P}}_3}{\hat{P}_3^2}+\cdots][1+\cdots] [(\lambda\psi^T)\frac{-\slashed{\hat{P}}_1}{\hat{P}_1^2}+\cdots]+\mathcal{P}(123)\notag\\
 =& i\lambda^3 \bar{\psi}\psi^T  \frac{\slashed{\hat{P}}_3}{\hat{P}_3^2} \frac{\slashed{\hat{P}}_1}{\hat{P}_1^2}+\cdots+\mathcal{P}(123)
\end{align}
where we just write the first order terms in the expansion of each
propagator. When $z\rightarrow \infty$ (see also \eref{arg-1}),
\begin{align}
  \frac{\slashed{\hat{P}}_3}{\hat{P}_3^2}\frac{\slashed{\hat{P}}_1}{\hat{P}_1^2}
=\frac{\hat{p}_3\cdot \hat{p}_1}{\hat{p}_3^2\hat{p}_1^2}
= &\frac{(\hat{p}_3+P_3)\cdot (\hat{p}_1+P_1)}{(\hat{p}_3+P_3)^2  (\hat{p}_1+P_1)^2} \notag\\
= & \frac{(p_3+P_3)\cdot(p_1+P_1)+z[q_3\cdot (p_1+P_1)+q_1\cdot (p_3+P_3)]}{[(p_3+P_3)^2+zq_3\cdot(p_3+P_3)][(p_1+P_1)^2+zq_1\cdot (p_1+P_1)]}
\sim  \frac{1}{z}
\end{align}
where $P_1,P_3$ appear because of the same reasons we have explained
in the previous section, i.e., with the  added  soft momenta. The
above result shows that the lieading term vanishs in the limit of
$z\rightarrow \infty$, and the next terms will also vanish because
they contain more derivatives as showed in \eref{con:D}.

For the second case $\langle
H_1\Psi_{2\alpha}\bar\Psi_{3\beta}\rangle$, the corresponding
Feynman diagrams are showed in Figure \ref{con:Yukawacaseonetwo}.
Since there are five contributing digrams,  the expression of
$\hat{G}_3$ is a little complicated
\begin{align}
 \hat{G}_3(z)&= \left[ (-i\mathcal{D}^{-1}_{23})(i\lambda)(-i\mathcal{D}^{-1}_{23})(-i\mathcal{D}^{-1}_{11})
 +(-i\mathcal{D}^{-1}_{23})(i\lambda)(-i\mathcal{D}^{-1}_{12})(-i\mathcal{D}^{-1}_{13})
 +(-i\mathcal{D}^{-1}_{13})(i\lambda)(-i\mathcal{D}^{-1}_{23})(-i\mathcal{D}^{-1}_{12}) \right. \notag\\
 &+\left. (-i\mathcal{D}^{-1}_{13})(i\lambda)(-i\mathcal{D}^{-1}_{22})(-i\mathcal{D}^{-1}_{13})
 +(-i\mathcal{D}^{-1}_{33})(i\lambda)(-i\mathcal{D}^{-1}_{12})(-i\mathcal{D}^{-1}_{12})\right] .
\end{align}
After LSZ reduction, we get the $z$-dependence of the three-deformed amplitude
\begin{align}
  i\widetilde{B}_3(z)
  =&i\lambda(\lim_{\hat{p}_1^2\rightarrow0}\hat{p}_1^2)(\lim_{\hat{p}_2^2\rightarrow0}\hat{\slashed{p}}_2) \bar{u}_{s_2}(\hat{p}_2)
  \left[ \mathcal{D}^{-1}_{23}(\hat{P}_2)\mathcal{D}^{-1}_{23}(\hat{P}_3)\mathcal{D}^{-1}_{11}(\hat{P}_1)
  +\mathcal{D}^{-1}_{23}(\hat{P}_2)\mathcal{D}^{-1}_{12}(\hat{P}_3)\mathcal{D}^{-1}_{13}(\hat{P}_1) \right. \notag\\
  +&\left. \mathcal{D}^{-1}_{13}(\hat{P}_2)\mathcal{D}^{-1}_{23}(\hat{P}_3)\mathcal{D}^{-1}_{12}(\hat{P}_1)
   +\mathcal{D}^{-1}_{13}(\hat{P}_1)\mathcal{D}^{-1}_{22}(\hat{P}_3)\mathcal{D}^{-1}_{13}(\hat{P}_2)
  +\mathcal{D}^{-1}_{33}(\hat{P}_2)\mathcal{D}^{-1}_{12}(\hat{P}_3)\mathcal{D}^{-1}_{12}(\hat{P}_1)\right] v_{s_3}(\hat{p}_3) \lim_{\hat{p}_3^2\rightarrow0}\hat{\slashed{p}}_3, \notag \label{yukawa-first}
\end{align}
where $s_2,s_3$ label the helicities of fermions. In the above
formula  the leading order is given the first term with two
$\mathcal{D}_{23}^{-1}$ as
\begin{align}
  (\lim_{\hat{p}_1^2\rightarrow0}\hat{p}_1^2)(\lim_{\hat{p}_2^2\rightarrow0}\hat{\slashed{p}}_2) \bar{u}_{s_2}(\hat{p}_2)
   \frac{\hat{\slashed{p}}_2^T}{\hat{p}_2^2} \frac{\hat{\slashed{p}}_3^T}{\hat{p}_3^2} \frac{1}{\hat{p}_1^2} v_{s_3}(\hat{p}_3) \lim_{\hat{p}_3^2\rightarrow0}\hat{\slashed{p}}_3 \sim  \bar{u}_{s_2}(\hat{p}_2) v_{s_3}(\hat{p}_3),
\end{align}
then the large $z$ behavior of $\widetilde{B}_3(z)$ depends on
external  wavefunction of fermions. The reason why the first term in
\eref{yukawa-first} contributes as the leading order is that its
three propagators are all in the first order in the expansion of
$D^{-1}$ with least derivatives. For example of four dimensions, if
we make a three-leg deformation like \eref{three-shift}, the
external wavefunctions are also deformed as showed in
\cite{Elvang:2013cua}, then the leading order is at least in the
order $O(z^0)$. So we conclude that $\widetilde{B}_3(z)$ always has
non-zero boundary contributions in four dimensions.\footnote{In
general dimensions, the deformation of external wavefunctions are
complicated, then we don't talk about here.}

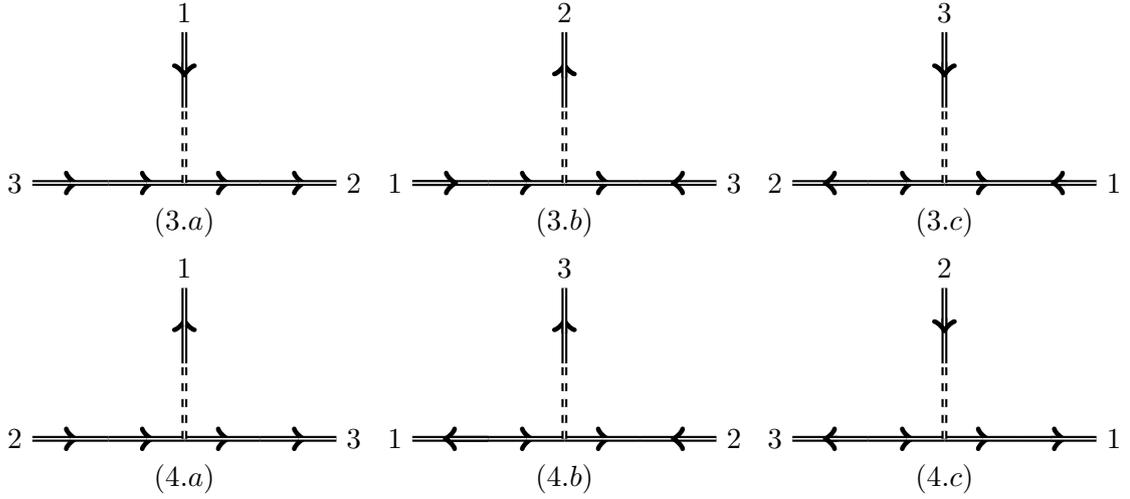
\begin{figure}
  \begin{center}
    \begin{tikzpicture}
      %%%%%%% diagrams for the second case
      \draw [double,thick,->] (0,0)--(0.6,0);
      \draw [double,thick] (0.5,0)--(1,0);
      \draw [double,thick,->] (1,0)--(1.6,0);
      \draw [double,thick] (1.5,0)--(2,0);
      \draw [double,thick,->] (-2,0)--(-1.4,0);
      \draw [double,thick] (-1.5,0)--(-1,0);
      \draw [double,thick,->] (-1,0)--(-0.4,0);
      \draw [double,thick] (-0.5,0)--(0,0);
      \draw [double,thick,dashed] (0,0)--(0,1);
      \draw [double,thick] (0,1)--(0,1.6);
      \draw [double,thick,<-] (0,1.4)--(0,2);
      \node [left] at (-2,0) {$3$};
      \node [above] at (0,2) {$1$};
      \node [right] at (2,0) {$2$};
      \node [below] at (0,-0.2) {$(3.a)$};
      %%%%%%
      \draw [double,thick,->] (5,0)--(5.6,0);
      \draw [double,thick] (5.5,0)--(6,0);
      \draw [double,thick] (6,0)--(6.5,0);
      \draw [double,thick,<-] (6.4,0)--(7,0);
      \draw [double,thick] (3,0)--(4,0);
      \draw [double,thick,->] (3,0)--(3.6,0);
      \draw [double,thick,->] (4,0)--(4.6,0);
      \draw [double,thick] (4.5,0)--(5,0);
      \draw [double,thick,dashed] (5,0)--(5,1);
      \draw [double,thick,->] (5,1)--(5,1.6);
      \draw [double,thick] (5,1.4)--(5,2);
      \node [left] at (3,0) {$1$};
      \node [above] at (5,2) {$2$};
      \node [right] at (7,0) {$3$};
      \node [below] at (5,-0.2) {$(3.b)$};
      %%%%%%%
     \draw [double,thick,->] (10,0)--(10.6,0);
     \draw [double,thick] (10.5,0)--(11,0);
     \draw [double,thick] (11,0)--(11.6,0);
     \draw [double,thick,<-] (11.4,0)--(12,0);
     \draw [double,thick] (8,0)--(8.6,0);
     \draw [double,thick,<-] (8.4,0)--(9,0);
     \draw [double,thick,->] (9,0)--(9.6,0);
     \draw [double,thick] (9.5,0)--(10,0);
     \draw [double,thick,dashed] (10,0)--(10,1);
     \draw [double,thick] (10,1)--(10,1.6);
     \draw [double,thick,<-] (10,1.4)--(10,2);
     \node [left] at (8,0) {$2$};
     \node [above] at (10,2) {$3$};
     \node [right] at (12,0) {$1$};
     \node [below] at (10,-0.2) {$(3.c)$};
    \end{tikzpicture}

    \begin{tikzpicture}
      %%%%%%% diagrams for the second case
      \draw [double,thick,->] (0,0)--(0.6,0);
      \draw [double,thick] (0.5,0)--(1,0);
      \draw [double,thick,->] (1,0)--(1.6,0);
      \draw [double,thick] (1.5,0)--(2,0);
      \draw [double,thick,->] (-2,0)--(-1.4,0);
      \draw [double,thick] (-1.5,0)--(-1,0);
      \draw [double,thick,->] (-1,0)--(-0.4,0);
      \draw [double,thick] (-0.5,0)--(0,0);
      \draw [double,thick,dashed] (0,0)--(0,1);
      \draw [double,thick,->] (0,1)--(0,1.6);
      \draw [double,thick] (0,1.5)--(0,2);
      \node [left] at (-2,0) {$2$};
      \node [above] at (0,2) {$1$};
      \node [right] at (2,0) {$3$};
      \node [below] at (0,-0.2) {$(4.a)$};
      %%%%%%
      \draw [double,thick,->] (5,0)--(5.6,0);
      \draw [double,thick] (5.5,0)--(6,0);
      \draw [double,thick] (6,0)--(6.5,0);
      \draw [double,thick,<-] (6.4,0)--(7,0);
      \draw [double,thick] (3,0)--(4,0);
      \draw [double,thick,<-] (3.4,0)--(4,0);
      \draw [double,thick,->] (4,0)--(4.6,0);
      \draw [double,thick] (4.5,0)--(5,0);
      \draw [double,thick,dashed] (5,0)--(5,1);
      \draw [double,thick,->] (5,1)--(5,1.6);
      \draw [double,thick] (5,1.4)--(5,2);
      \node [left] at (3,0) {$1$};
      \node [above] at (5,2) {$3$};
      \node [right] at (7,0) {$2$};
      \node [below] at (5,-0.2) {$(4.b)$};
      %%%%%%%
     \draw [double,thick,->] (10,0)--(10.6,0);
     \draw [double,thick] (10.5,0)--(11,0);
     \draw [double,thick,->] (11,0)--(11.6,0);
     \draw [double,thick] (11.4,0)--(12,0);
     \draw [double,thick] (8,0)--(8.6,0);
     \draw [double,thick,<-] (8.4,0)--(9,0);
     \draw [double,thick,->] (9,0)--(9.6,0);
     \draw [double,thick] (9.5,0)--(10,0);
     \draw [double,thick,dashed] (10,0)--(10,1);
     \draw [double,thick] (10,1)--(10,1.6);
     \draw [double,thick,<-] (10,1.4)--(10,2);
     \node [left] at (8,0) {$3$};
     \node [above] at (10,2) {$2$};
     \node [right] at (12,0) {$1$};
     \node [below] at (10,-0.2) {$(4.c)$};
  \end{tikzpicture}
  \end{center}
  \caption{The three Feynman diagrams in the first line are for $\langle \Psi_{1\alpha}\bar\Psi_{2\beta}\Psi_{3\gamma}\rangle$, the three Feynman diagrams in the second line are for $\langle \bar\Psi_{1\alpha}\Psi_{2\beta}\bar\Psi_{3\gamma}\rangle$.} \label{con:Yukawacase34}
\end{figure}

For the third case $\langle
\Psi_{1\alpha}\bar\Psi_{2\beta}\Psi_{3\gamma}\rangle$, there are
three Feynman diagrams contributing as showed in Figure
\ref{con:Yukawacase34}. The deformed correlation function
$\hat{G}_3$ is
\begin{align}
\hat{G}_3(z)
=& (-i\mathcal{D}_{13}^{-1})(i\lambda)(-i\mathcal{D}_{22}^{-1})(-i\mathcal{D}_{23}^{-1})
+ (-i\mathcal{D}_{33}^{-1})(i\lambda(-i\mathcal{D}_{22}^{-1})(-i\mathcal{D}_{12}^{-1}) \notag\\
+& (-i\mathcal{D}_{23}^{-1})(i\lambda)(-i\mathcal{D}_{23}^{-1})(-i\mathcal{D}_{12}^{-1})
\end{align}
After LSZ reduction,
\begin{align}
  i\widetilde{B}_3(z)&=(i\lim_{\hat{p}_1^2\rightarrow0}\hat{\slashed{p}}_1)\bar{u}_{s_1}(\hat{p}_1)(i\lim_{\hat{p}_3^2\rightarrow0}\hat{\slashed{p}}_3) \bar{u}_{s_3}(\hat{p}_3) \hat{G}_3(z) (i\lim_{\hat{p}_2^2\rightarrow0}\hat{\slashed{p}}_2) v_{s_2}(\hat{p}_2),
\end{align}
just like in the previous case we only need to focus on the third
term and analyze the large $z$ behavior of it, since the term
contains the propagator $\mathcal{D}_{12}^{-1}$ which contributes as
$O(z)$ after LSZ reduction. So when considering the contributions
from wavefunctions, in some helicity configurations  the
$\widetilde{B}_3(z)$ vanishes, but also in some other helicity
configurations  non-zero boundary contributions appear. As for the
last case $\langle
\bar\Psi_{1\alpha}\Psi_{2\beta}\bar\Psi_{3\gamma}\rangle$, whose
Feynman diagrams are shown in Figure \eref{con:Yukawacase34}, the
discussions are same as for the third case, since both are related
by the complex conjugation.

\section{Yang-Mills Theory} \label{YM}

In this section we apply the same method to the Yang-Mills theory.
In  \cite{Cheung:2008dn}, the author used the background field
method by splitting the YM field into a hard field and a soft field,
but only wrote down quadratic terms of hard field since they only
considered the significance of propagators of hard field. However,
because we consider the on-shell recursion relations of three legs
deformed, we need to calculate interaction part of hard field. As
before, we consider the Lagrangian of the pure Yang-Mills theory
\begin{equation}
\mathcal{L}=-\frac{1}{4}F_{\mu\nu}^cF^{\mu\nu c}.
\end{equation}
If we split the Yang-Mills field $A_{\mu}$ into $A_{\mu}\rightarrow a_{\mu}+A_{\mu}$,
where $a_{\mu}$ is the soft field and $A_{\mu}$ is the hard field, the field strength becomes
\begin{align}
F_{\mu\nu}^c\rightarrow \bar{F}_{\mu\nu}^c+(\bar{\mathcal{D}}_{\mu}A_{\nu})^c-(\bar{\mathcal{D}}_{\nu}A_{\mu})^c+g f^{abc}A_{\mu}^aA_{\nu}^b,
\end{align}
where $\bar{F}_{\mu\nu}^c$ is the field strength for soft field and $\bar{\mathcal{D}}_{\mu}^{ab}=\partial_{\mu}\delta^{ab}-g f^{cab}a^c_{\mu}$ is the background covariant derivative.
Substituting it into the Lagrangian, we obtain
\begin{align}
\mathcal{L}(a,A)=& -\frac{1}{2}(\bar{\mathcal{D}}_{\mu}A_{\nu})^a (\bar{\mathcal{D}}^{\mu}A^{\nu})^a+\frac{1}{2}(\bar{\mathcal{D}}_{\mu}A_{\nu})^a (\bar{\mathcal{D}}^{\nu}A^{\mu})^a
-\frac{1}{2}gf^{abc}\bar{F}^{c\mu\nu}A^{a}_{\mu}A^{b}_{\nu} \notag \\
&-gf^{abc}A^{a\mu}A^{b\nu}(\mathcal{\bar D}_{\mu}A_{\nu})^c
-\frac{1}{4}g^2f^{abe}f^{cde}A^{a\mu}A^{b\nu}A^c_{\mu}A^d_{\nu},
\end{align}
where these terms without $A$ or linear in $A$ have been dropped.
Then we consider adding the gauge-fixing term,\footnote{We don't consider the ghost term for our considerations are limited in the tree diagrams.}
\begin{equation}\label{bggf}
\mathcal{L}_{gf}=-\frac{1}{2}{\xi^{-1}}(\mathcal{\bar D}^{\mu}A_{\mu})^a(\mathcal{\bar D}^{\nu}A_{\nu})^a
=-\frac{1}{2}(\mathcal{\bar D}^{\mu}A^{\nu})^c(\mathcal{\bar D}_{\nu}A_{\mu})^c-\frac{1}{2}gf^{abc}\bar{F}^{c\mu\nu}A^{a}_{\mu}A^{b}_{\nu}
\end{equation}
with $\xi=1$. So we get
\begin{align}
  \mathcal{L}(a,A)+\mathcal{L}_{gf}
  =&-\frac{1}{2}(\bar{\mathcal{D}}_{\mu}A_{\nu})^a (\bar{\mathcal{D}}^{\mu}A^{\nu})^a
  -gf^{abc}\bar{F}^{c\mu\nu}A^{a}_{\mu}A^{b}_{\nu} \notag \\
  &-gf^{abc}A^{a\mu}A^{b\nu}(\mathcal{\bar D}_{\mu}A_{\nu})^c
  -\frac{1}{4}g^2f^{abe}f^{cde}A^{a\mu}A^{b\nu}A^c_{\mu}A^d_{\nu} \label{YM-Lagrangian},
\end{align}
then we can get the expression of propagator from the first line and vertices from the second line.

First, to write down the expression of propagator, we need to consider the first line of \eref{YM-Lagrangian} and after using integration by parts we get
\begin{align}
M^{ab}_{\mu\nu}&=(\mathcal{\bar D}^{\rho}\mathcal{\bar D}_{\rho})^{ab}g_{\mu\nu}-2gf^{abc}\bar{F}^c_{\mu\nu} \notag \\
&= g_{\mu\nu}\delta^{ab}\partial^2+gf^{abc}g_{\mu\nu}(\partial^{\rho}a_{\rho}^c)+2gf^{abc}g_{\mu\nu} a_{\rho}^c\partial^{\rho} - 2gf^{abc}\bar{F}_{\mu\nu}^c
+g^2f^{adc}f^{ceb} {a_{\rho}^d}a^{\rho e} g_{\mu\nu} \notag \\
&=g_{\mu\nu}\delta^{ab}\partial^2 - V^{ab}_{\mu\nu}
\end{align}
where $V^{ab}_{\mu\nu} = -gf^{abc}g_{\mu\nu}(\partial^{\rho}a_{\rho}^c)-2gf^{abc}g_{\mu\nu} a_{\rho}^c\partial^{\rho} + 2gf^{abc}\bar{F}_{\mu\nu}^c
-g^2f^{adc}f^{ceb}{a_{\rho}^d}a^{\rho e}g_{\mu\nu}$.
Then we take the inverse of $M_{\mu\nu}^{ab}$ formlly as
\begin{align}
 (M^{ab}_{\mu\nu})^{-1}
 =& (g_{\mu\nu}\delta^{ab}\partial^2 - V^{ab}_{\mu\nu})^{-1} \notag\\
 =& \left\{\partial^2 \left[g_{\mu\nu}\delta^{ab}-(\partial^2)^{-1}  V^{ab}_{\mu\nu}\right] \right\}^{-1} \notag\\
 =& \left[g^{\mu\nu}\delta^{ab} -(\partial^2)^{-1}  V^{ab,\mu\nu}+(\partial^2)^{-1} V^{ac,\mu\rho}(\partial^2)^{-1}  V^{cb,\rho\nu} +\cdots \right] (\partial^2)^{-1},
\end{align}
the momentum is got by replacing $\partial^2,\partial_{\mu}$ by
$-p^2,i p_{\mu}$. Second, from \eref{YM-Lagrangian}, we know there
are two kinds  of cubic vertex and one quartic vertex, then we need
to write the explicit expressions of these vertices one by one. The
first cubic vertex contains three hard momenta only and is given by
\begin{align}
iV^{abc}_{\mu\nu\rho}(p,q,r)=&gf^{abc}\left[(q-r)_{\mu}g_{\nu\rho}+(r-p)_{\nu}g_{\rho\mu}+(p-q)_{\rho}g_{\mu\nu}\right].
\end{align}
The second cubic vertex connects three hard fields and one soft
field
\begin{align}
iV^{abc}_{\mu\nu\rho}(a)=&-ig^2\left[f^{abd}f^{dec}(a^{e}_{\mu}g_{\nu\rho}-a^{e}_{\nu}g_{\mu\rho}) +f^{acd}f^{deb}(a^e_{\mu}g_{\nu\rho}-a^e_{\rho}g_{\mu\nu}) +f^{bcd}f^{dea}(a^e_{\nu}g_{\mu\rho}-a^e_{\rho}g_{\mu\nu})   \right].
\end{align}
Although we will not use the quartic vertex
$iV^{abcd}_{\mu\nu\rho\sigma}$ of  hard fields in this paper, when
four or more momenta are deformed, the quartic vertex should be
included to analyze the boundary behavior, so we give its expression
\begin{align}
iV^{abcd}_{\mu\nu\rho\sigma} =
-i g^2 \left[ f^{abe}f^{cde} (g_{\mu\rho}g_{\nu\sigma}-g_{\mu\sigma}g_{\nu\rho}) +f^{ade}f^{bce} (g_{\mu\nu}g_{\sigma\rho}-g_{\mu\rho}g_{\sigma\nu})+f^{ace}f^{dbe} (g_{\mu\sigma}g_{\rho\nu}-g_{\mu\nu}g_{\rho\sigma}) \right]\notag
\end{align}
for completeness.
\\

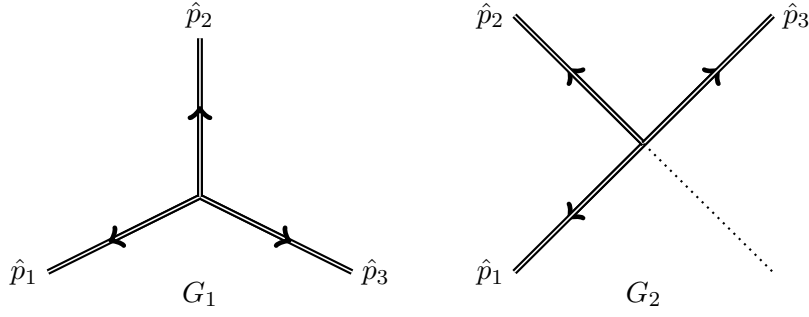
\begin{figure}
  \begin{center}
    \begin{tikzpicture}
      %%%%%%% diagrams for the second case
      \draw [double,thick] (0,0)--(2,-1);
      \draw [double,thick,->] (0,0)--(1.2,-0.6);
      \draw [double,thick] (-2,-1)--(0,0);
      \draw [double,thick,<-] (-1.2,-0.6)--(0,0);
      \draw [double,thick] (0,0)--(0,2.1);
      \draw [double,thick,->] (0,0)--(0,1.2);
      \node [left] at (-2,-1) {$\hat{p}_1$};
      \node [above] at (0,2.1) {$\hat{p}_2$};
      \node [right] at (2,-1) {$\hat{p}_3$};
      \node [below] at (0,-1) {$G_1$};
    \end{tikzpicture} \qquad
    \begin{tikzpicture}
      %%%%%%
      \draw [double,thick,->] (0,0)--(-1,-1);
      \draw [double,thick] (0,0)--(-1.7,-1.7);
      \draw [double,thick,->] (0,0)--(-1,1);
      \draw [double,thick] (0,0)--(-1.7,1.7);
      \draw [double,thick,->] (0,0)--(1,1);
      \draw [double,thick] (0,0)--(1.7,1.7);
      \draw [thick,dotted] (0,0)--(1.7,-1.7);
      \node [left] at (-1.7,-1.7) {$\hat{p}_1$};
      \node [left] at (-1.7,1.7) {$\hat{p}_2$};
      \node [right] at (1.7,1.7) {$\hat{p}_3$};
      \node [below] at (0,-1.7) {$G_2$};
      %%%%%%%
    \end{tikzpicture}
  \end{center}
  \caption{The Feynman diagrams with three hard lines for deformed correlation function in Yang-Mills theory} \label{diagrams-YM}
\end{figure}

Having obtained the Feynman rules,  we begin to  calculate deformed
correlation function of an amplitude. Since we only deform three
external momenta,  we need only consider these diagrams with three
hard particles. So there are only two diagrams contributing as shown
in Figure \ref{diagrams-YM}, the first is the one given by the first
kind of cubic vertex, denoted by $G_1$, the second is the one given
by the second kind of cubic vertex denoted by $G_2$.
The expressions of two diagrams are given as
\begin{align}
\hat{G}_3^{def,\alpha\beta\gamma}(z)&=iV^{abc}_{\mu\nu\rho} \left[-i(M^{-1})^{ad,\mu\alpha}(\hat{p}_1)\right]  \left[-i(M^{-1})^{be,\nu\beta}(\hat{p}_2)\right]  \left[-i(M^{-1})^{cf,\rho\gamma}(\hat{p}_3)\right] \notag\\
&+iV^{abc}_{\mu\nu\rho}(a) \left[-i(M^{-1})^{ad,\mu\alpha}(\hat{p}_1)\right]  \left[-i(M^{-1})^{be,\nu\beta}(\hat{p}_2)\right]  \left[-i(M^{-1})^{cf,\rho\gamma}(\hat{p}_3)\right]. \label{YM-correlation}
\end{align}
Then we do the LSZ reduction and get the boundary contribution of
the  three-leg deformed amplitude,
\begin{align}
 \widetilde{B}_3(z)= (i\lim_{\hat{p}_1^2\rightarrow0}\hat{p}_1^2)(i\lim_{\hat{p}_2^2\rightarrow0}\hat{p}_2^2)(i\lim_{\hat{p}_3^2\rightarrow0}\hat{p}_3^2) \hat{G}_3^{def,\alpha\beta\gamma}(z) \epsilon_{s_1}^{\alpha}(\hat{p}_1) \epsilon_{s_2}^{\beta}(\hat{p}_2) \epsilon_{s_3}^{\gamma}(\hat{p}_3),
\end{align}
where $s_i$ represents the helicity of the $i$th particle. Since
there are two diagrams contributing, we discuss  them one by one.
The first contribution comes from the first line of
\eref{YM-correlation},
\begin{align}
  &\widetilde{B}^{(1)}_3(z) \notag\\
  =&(\lim_{\hat{p}_1^2\rightarrow0}\hat{p}_1^2)(\lim_{\hat{p}_2^2\rightarrow0}\hat{p}_2^2)(\lim_{\hat{p}_3^2\rightarrow0}\hat{p}_3^2)
  V^{abc}_{\mu\nu\rho} \left[(M^{-1})^{ad,\mu\alpha}(\hat{p}_1)\right]  \left[(M^{-1})^{be,\nu\beta}(\hat{p}_2)\right]  \left[(M^{-1})^{cf,\rho\gamma}(\hat{p}_3)\right] \epsilon_{s_1}^{\alpha}(\hat{p}_1) \epsilon_{s_2}^{\beta}(\hat{p}_2) \epsilon_{s_3}^{\gamma}(\hat{p}_3) \notag\\
  =& gf^{abc} \left[(\hat{p}_2-\hat{p}_3)_{\mu}g_{\nu\rho}+(\hat{p}_3-\hat{p}_1)_{\nu}g_{\rho\mu}+(\hat{p}_1-\hat{p}_2)_{\rho}g_{\mu\nu} \right] \epsilon_{s_1}^{\alpha}(\hat{p}_1) \epsilon_{s_2}^{\beta}(\hat{p}_2) \epsilon_{s_3}^{\gamma}(\hat{p}_3) \notag\\
  &\left[g^{\mu\alpha}\delta^{ad}-(\hat{p}_1^2)^{-1}V^{ad,\mu\alpha}+\cdots \right]
  \left[g^{\nu\beta}\delta^{be}-(\hat{p}_2^2)^{-1}V^{be,\nu\beta}+\cdots \right]
  \left[g^{\rho\gamma}\delta^{cf}-(\hat{p}_1^2)^{-1}V^{cf,\rho\gamma}+\cdots \right]. \label{YM-firstpart}
\end{align}
If we expand the above equation and look at the first order term, we
obtain
\begin{align}
 f^{def} \left[(\hat{p}_2-\hat{p}_3)\cdot \epsilon_{s_1}(\hat{p}_1)\right] \left[ \epsilon_{s_2}(\hat{p}_2)\cdot \epsilon_{s_3}(\hat{p}_3) \right] + \mathcal{P}(123).
\end{align}
For the higher order terms, although $V^{ad,\mu\alpha}$ may
contribute in the order $z$,  $(\hat{p}_2^2)^{-1}$ is also $z^{-1}$,
so higher order terms contribute same order as the first ones. So
the large $z$ behavior of $\hat{A}^{(1)}_3(z)$ is more complicated,
which depends on the $z$ behaviors of polarization vectors as well
as  three deformed momenta. Contribution from the second diagram is
\begin{align}
  &\widetilde{B}^{(2)}_3(z) \notag\\
  =& (\lim_{\hat{p}_1^2\rightarrow0}\hat{p}_1^2)(\lim_{\hat{p}_2^2\rightarrow0}\hat{p}_2^2)(\lim_{\hat{p}_3^2\rightarrow0}\hat{p}_3^2)
  V^{abc}_{\mu\nu\rho}(a) \left[(M^{-1})^{ad,\mu\alpha}(\hat{p}_1)\right]  \left[(M^{-1})^{be,\nu\beta}(\hat{p}_2)\right]  \left[(M^{-1})^{cf,\rho\gamma}(\hat{p}_3)\right]  \epsilon_{s_1}^{\alpha}(\hat{p}_1) \epsilon_{s_2}^{\beta}(\hat{p}_2) \epsilon_{s_3}^{\gamma}(\hat{p}_3) \notag\\
  =& -g^2\left[f^{abc}f^{ced}(a^{e}_{\mu}g_{\nu\rho}-a^{e}_{\nu}g_{\mu\rho}) +f^{adc}f^{ceb}(a^e_{\mu}g_{\nu\rho}-a^e_{\rho}g_{\mu\nu}) +f^{bdc}f^{cea}(a^e_{\nu}g_{\mu\rho}-a^e_{\rho}g_{\mu\nu}) \epsilon_{s_1}^{\alpha}(\hat{p}_1) \epsilon_{s_2}^{\beta}(\hat{p}_2) \epsilon_{s_3}^{\gamma}(\hat{p}_3) \right] \notag\\
  &\left[g^{\mu\alpha}\delta^{ad}-(\hat{p}_1^2)^{-1}V^{ad,\mu\alpha}+\cdots \right] \left[g^{\nu\beta}\delta^{be}-(\hat{p}_2^2)^{-1} V^{be,\nu\beta}+\cdots \right] \left[g^{\rho\gamma}\delta^{cf}-(\hat{p}_1^2)^{-1}V^{cf,\rho\gamma}+\cdots \right], \label{YM-secondpart}
\end{align}
where  terms of the first order are
\begin{align}
 f^{def}f^{fed}\{ \left[a\cdot\epsilon_{s_1}(\hat{p}_1)\right] \left[\epsilon_{s_2}(\hat{p}_2)\cdot \epsilon_{s_3}(\hat{p}_3) \right]-\left[a\cdot\epsilon_{s_2}(\hat{p}_2)\right] \left[\epsilon_{s_1}(\hat{p}_1)\cdot \epsilon_{s_3}(\hat{p}_3)\right] \},
\end{align}
and higher order terms behave similarly as before.
% so the large $z$ behavior of $\hat{A}^{(2)}_3(z)$ solely depends on the behaviors of polarization vectors.
Combing the two contributions, since the first contribution depends
on deformed momenta and always have larger order of $z$ than the
second one,  when we analyze the large $z$ behavior of amplitude, we
can only consider the first diagram and ignore the second one.
\\

After calculating the boundary terms, to specify the explicit
boundary  behavior of amplitude, we should know the $z$-dependences
of polarization vectors. Since the $z$-dependence of polarization
vectors in three-leg deformed (or even more deformed) case are not
so easily determined,  we take an amplitude in four dimensions as an
example. The three-leg deformation is chosen as\footnote{We can
easily check that they satisfy the previous mentioned conditions.
And here the conventions follow \cite{Elvang:2013cua}.}
 \begin{align}
|\hat{1}\rangle &= |1\rangle -z|2\rangle -z|3\rangle, \quad |\hat{1}]=|1],  \notag\\
|\hat{2}] &= |2] +z|1], \quad |\hat{2}\rangle=|2\rangle, \notag\\
|\hat{3}] &= |3] +z|1], \quad |\hat{3}\rangle=|3\rangle. \label{three-shift}
\end{align}
Next we write down the corresponding polarization vectors
\begin{align}
\epsilon_1^-(z) &= -\frac{\spab{\hat{1}| \gamma^{\mu} |q_1}}{\sqrt{2} \spbb{q_1~\hat{1}}} \quad    \epsilon_1^+(z) = -\frac{\spab{q_1| \gamma^{\mu} |\hat{1}}}{\sqrt{2} \spaa{q_1~\hat{1}}},    \notag\\
\epsilon_2^-(z) &= -\frac{\spab{\hat{2}| \gamma^{\mu} |q_2}}{\sqrt{2} \spbb{q_2~\hat{2}}} \quad    \epsilon_2^+(z) = -\frac{\spab{q_2| \gamma^{\mu} |\hat{2}}}{\sqrt{2} \spaa{q_2~\hat{2}}},   \notag\\
\epsilon_3^-(z) &= -\frac{\spab{\hat{3}| \gamma^{\mu} |q_3}}{\sqrt{2} \spbb{q_3~\hat{3}}} \quad    \epsilon_3^+(z) = -\frac{\spab{q_3| \gamma^{\mu} |\hat{3}}}{\sqrt{2} \spaa{q_3~\hat{3}}}.  \label{three-epsilon}
\end{align}
From the above form of the polarization vectors, we can know that the polarization vectors $\epsilon_1^-(z), \epsilon_2^+(z),  \epsilon_3^+(z) $ are of order $\mathcal{O}(z)$, while the polarization vectors $\epsilon_1^+(z), \epsilon_2^-(z), \epsilon_3^-(z) $ are of order $O(z^{-1})$. There are total $8$ helicity configurations for the three particles, since the analysis for each case are same, then we focus on one case $(1^+,2^-,3^-)$ to illustrate our method. When $z\rightarrow \infty$, the leading contributions for the amplitude of the helicity configuration are the leading terms in the first diagram as
\begin{align}
  \left[ (\hat{p}_2-\hat{p}_3) \cdot \epsilon_1^+(z)\right]\left[\epsilon_2^-(z)\cdot \epsilon_3^-(z)\right] + \left[ (\hat{p}_3-\hat{p}_1) \cdot \epsilon_2^-(z)\right]\left[\epsilon_3^-(z)\cdot \epsilon_1^+(z)\right] + \left[ (\hat{p}_1-\hat{p}_2) \cdot \epsilon_3^-(z)\right]\left[\epsilon_1^+(z)\cdot \epsilon_2^-(z)\right].  \notag
\end{align}
If we choose $q_2=q_3$, then the first term in the above formula vanish for $\left[\epsilon_2^-(z)\cdot \epsilon_3^-(z)\right]=0$, and after calculations we can know that $\left[\epsilon_3^-(z)\cdot \epsilon_1^+(z)\right]\sim 1/z^2, \left[\epsilon_1^+(z)\cdot \epsilon_2^-(z)\right] \sim 1/z^2$ and $\left[ (\hat{p}_3-\hat{p}_1) \cdot \epsilon_2^-(z)\right]\sim z, \left[ (\hat{p}_1-\hat{p}_2) \cdot \epsilon_3^-(z)\right]\sim z$. So when $z\rightarrow \infty$, the boundary terms vanish for gluon amplitude with $(1^+,2^-,3^-)$ for three deformed particles. However we should note that for other helicity configurations, the boundary terms are not always vanish, which can be inferred from the $z$ dependences of $\epsilon_1^-(z), \epsilon_2^+(z),  \epsilon_3^+(z) $.

\section{Conclusion} \label{conclusion}
The vanishing of boundary contribution $B_n$ of a deformed amplitude
$\hat{A}_n$ is vital for the existence of on-shell recursion
relations, so many literatures have paid special attentions  to
analyze how a deformed amplitude $\hat{A}_n$ behaves when $z$
approaches infinity and to determine when these recursion relations
are applicable. However, in general the boundary contribution is
unavoidable in many theories, so the understanding of boundary
becomes an interesting problem.

In this paper, we try to calculate the $z$-denpendence of a deformed
amplitude by using the background field method in the case of
multiple legs being deformed. The method relies on the key idea
proposed in \cite{ArkaniHamed:2008yf,Cheung:2008dn} that we can view
the particles with momenta being deformed as hard particles while
others as soft particles when $z\rightarrow \infty$, and the
deformed amplitude is the description of hard particles scattering
in the background of soft particles. To apply the interpretation to
practical calculations, we need another tool, LSZ reduction, to
relate the deformed amplitude with its corresponding deformed
correlation function in \cite{Jin:2015pua}, where hard particles
correspond to fields with deformed momenta. Once the correspandance
is established, the computation of the $z$-dependence of the
deformed amplitude is transformed into the simple computation of the
sum of several Feynman diagrams, which exactly depict hard particles
scattering in the soft background. The key point is to write down
the correct Feynman rules for the hard Lagrangian, draw the
corresponding Feynman diagrams and calculate the expressions of
them. The whole procedure is the same as what we learn from the
standard QFT textbook, so the method is very simple. Although in
this paper we are limited in the case with only three external legs
being deformed,  it can also be generalized to  the situations where
more legs are deformed.

After given the general discussions of how to combine the background
field  method with on-shel recursion relations, we presented three
examples to illustrate our method from the simplest one, real scalar
theory, to more complicated  Yukawa theory and pure YM theory. There
are  some facts from these examples. First, the Feynman diagrams in
the background field method are actually the combination of all
diagrams which depict the real scattering process without
background, so  it's possible for us to get the explicit expressions
of the $z$-dependent part of deformed amplitude.  When we expand the
propagator in soft background, we can find these diagrams depicting
real scattering process again. Secondly, the large $z$ contributions
are dominated by the leading terms of the expansion of the
propagators in soft background, especially those with less
derivatives, because these derivatives will produce some
$\hat{p}_i^2\sim z$ in denominators. Thirdly, the external wave
functions are also important and actually determine if the on-shell
recursion relations exist in some critical cases, just as shown by
those examples in Yukawa theory and YM theory.

\section*{Acknowledgments}
We thank Bo Feng for giving us the idea and helpful discussions. We would also like to thank Qingjun Jing, Rijun Huang, Junjie Rao for helpful discussions. This work is support by Qiu-Shi Funding and the
National Natural Science Foundation of China (NSFC) with Grant No.11935013, No.11575156..

%%%%%%%%%%
\appendix

\bibliographystyle{JHEP}
\bibliography{reference}

\end{document}